# Valley Anisotropy in Elastic Metamaterials

Shuaifeng Li[1,2], Ingi Kim[3], Satoshi Iwamoto[3], Jianfeng Zang[2*], Jinkyu Yang[1*]

[1]Department of Aeronautics and Astronautics, University of Washington, Seattle, Washington, 98195-2400, USA

[2]School of Optical and Electronic Information and Wuhan National Laboratory for Optoelectronics, Huazhong University of Science and Technology, Wuhan, Hubei, 430074, China

[3]Institute of Industrial Science, University of Tokyo, Meguro, Tokyo, 153-8505, Japan

**Abstract**

Valley, as a new degree of freedom, raises the valleytronics in fundamental and applied science. The elastic analogs of valley states have been proposed by mimicking the symmetrical structure of either two-dimensional materials or photonic valley crystals. However, the asymmetrical valley construction remains unfulfilled. Here, we present the valley anisotropy by introducing asymmetrical design into elastic metamaterials. The elastic valley metamaterials are composed of bio-inspired hard spirals and soft materials. The anisotropic topological nature of valley is verified by asymmetrical distribution of the Berry curvature. We show the high tunability of the Berry curvature both in magnitude and sign enabled by our anisotropic valley metamaterials. Finally, we demonstrate the creation of valley topological insulators and show topologically protected propagation of transverse elastic waves relying on operating frequency. The proposed topological properties of elastic valley metamaterials pave the way to better understanding the valley topology and to creating a new type of topological insulators enabled by an additional valley degree of freedom.

**Introduction**

Elastic waves, possessing plenty of degree of freedoms (DOFs) including frequency, phase and polarization, have demonstrated tremendous promise in a variety of applications including target detection, information processing and biomedical imaging[1–4]. Recently, topology has been proposed as a new DOF in manipulating waves in both photonic and phononic systems, exhibiting remarkable impact not only on fundamental science such as condensed matter physics, but also on engineering applications such as low loss devices and waveguides[5–9]. In photonics, the photonic spin Hall effect has been achieved by taking advantage of spin DOF, which opens up an avenue of spin-dependent light transport and one-way spin transport[10–13]. In phononics, mechanical patterns and deformation have been employed as a new DOF to enable the elastic topological states[8,9,14–19].

Recently, valley, the degenerate yet inequivalent energy extrema in momentum space, has emerged as a new dimension in manipulating waves in electronics, photonics and phononics[5,6,20–24]. In graphene and transition metal dichalcogenides, valley-selective circular dichroism and valley Hall effect due to the long lifetime of valley polarization and non-zero Berry curvature have been studied for the promising applications in information carrier and storage[20,21,23,24]. As the concept of valley is introduced into the classic system, the photonic and phononic valley crystals are proposed, showing potential applications such as information processing via valley-dependent transportation[5,6,22]. However, existing designs of valley metamaterials are limited to the inherent spatial inversion symmetry of the physical system, where the typical Berry curvature distribution in the Brillouin zone follows $\Omega(-\boldsymbol{k}) = \Omega(\boldsymbol{k})$. The valley metamaterials without spatial inversion symmetry have not yet been explored. The introduction of asymmetrical design into valley crystals may add an additional degree of freedom in manipulating waves for waveguiding and information carrying purposes.

Here, we propose a new concept of valley anisotropy by introducing asymmetrical architecture into elastic metamaterials. The elastic valley metamaterials are composed of bio-inspired hard spiral scatterers and soft material matrix. The spiral structure ensures the system without spatial inversion symmetry. The valley anisotropy is revealed by the exceptional Berry curvature of this chiral anisotropic system. We show that the Berry curvature can be tuned by adjusting the parameters of spiral scatterers. By leveraging this enhanced design freedom, we demonstrate that our asymmetrical structure design with the integration of both soft and hard materials enables unprecedented topological manipulation of transverse elastic waves, allowing bending and stoppage of energy flow. Such manipulation of transverse waves can be useful in high-resolution imaging, such as trans-skull measurement and treatment in biomedical systems because of their high penetration and high contrast characteristics in human body[1,3,25]. The elastic valley states, carrying notable features of vortices, may also inspire new appealing applications in engineering.

**Results**

**Design of elastic valley metamaterials.** As illustrated in Fig. 1a, the elastic valley metamaterials are designed in a triangular lattice using the hard spirals as scatterers and the soft hydrogel as a matrix. Hydrogels are chosen as the soft matrix because they are acoustically similar to water and are ideally biocompatible materials[26]. The hard-spiral structure is inspired by the pattern on seashells. The spiral element is in the low order of symmetry falling into the point group $C_s$ that only contains identity and $\sigma_{xy}$ symmetries, which ensures the asymmetrical elastic valley metamaterial[27]. As a typical example in the monofilar spiral, Archimedean spiral structure is employed in our design of the unit cell. The key parameters are marked in Fig. 1a and detailed in Methods.

Fig. 1b and 1c present the band structure of the first Brillouin zone with the second and the third transverse wave bands separated by an omnidirectional band gap. Different from the Dirac dispersion in lattices with $C_{3v}$ symmetry, the band gap in our spiral system is a result of breaking the symmetry between the lattice and the scatterers, which is guaranteed by the asymmetric spiral structure in our design. The phononic band structure is a typical valley structure, which is similar to the band structure of transition metal dichalcogenides. Because of the chiral structure, we use six symbols from $K_1$ to $K_6$ to present the corners of the Brillouin zone. We illustrate the two bands along the edge of the first Brillouin zone (i.e., from $K_1$ to $K_6$) in Fig. 1c. We find that the both bands between $K_1$ and $K_2$ valleys are different from those between $K_2$ and $K_3$ valleys, and also between $K_3$ and $K_4$. This indicates that there exist three inequivalent valleys in our spiral structure. These three inequivalent valleys suggest the anisotropic band structure of the elastic valleys.

We further investigate the elastic valley states of the spiral structure in the eigen displacement field and valley polarizations. We choose $p$ and $q$ of the second and third bands at $K_1$ valley as two representing elastic valley states, as marked in Fig. 1c. Similar with electronic valley states, as well as the photonic and acoustic valley states, elastic valley states also exhibit a notable chirality, as shown in the eigen displacement field of $p$ and $q$ states in Fig. 1d. Fig. 1e illustrates the normalized pseudospin angular momentum density of phonons in $p$ and $q$ states, which is defined as $\boldsymbol{S} = \frac{\rho\omega}{2}\langle \boldsymbol{u}|\hat{\boldsymbol{S}}|\boldsymbol{u}\rangle$[28], where $\rho$ is the mass density, $\omega$ is the angular frequency and $\hat{\boldsymbol{S}}$ is the spin-1 operator. It can be simplified as $\boldsymbol{S} = \frac{\rho}{2\omega}Im(\boldsymbol{v}^*\times\boldsymbol{v})$, which indicates the rotation of velocity field. The local particle velocity rotates clockwise or counterclockwise driven by the phase difference. For convenience, we define the clockwise and the counterclockwise as pseudospin down and pseudospin up based on the right-hand rule. The pseudospin angular momentum characterizes the polarizations of phonons, e.g., linear, elliptical and circular polarizations. In the analysis of valley polarization, we find $p$ and $q$ state exhibit different polarizations. From Figs. 1d and e, both the amplitudes of displacement and the normalized pseudospin angular momentum reach maximum at the corner of the hexagon, indicating the circular polarizations at the corners. For example, when we investigate the *p* state, we find that the maximum of displacement field is located at three corners of hexagon. At the same time, the pseudospin angular momentum density mainly concentrates on the same places of the hexagon. Moreover, the negative value suggests that *p* state is in the pseudospin down state and the local particle velocity rotates clockwise. In stark contrast, *q* state is in the pseudospin-up state and local particle velocity rotates counterclockwise. It is interesting to note that when the *p* state exhibits circular (linear or elliptical) polarization at the three corners of hexagon where the *q* state exhibits linear or elliptical (circular) polarization.

The field distribution and normalized pseudospin angular momentum are completely different from the existing valley states in classic systems, where the normalized pseudospin angular momentum reaches maximum at the position where the amplitude of field becomes zero. The comparison of traditional acoustic valley vortex state is shown in supplementary figure 1 and supplementary note 1. The observed anomalous valley states have never been reported in the classic systems, which enrich the intrinsic physics of valley states and inspire the pseudospin source for communication.

**Deviated Berry curvature of the elastic valley metamaterials.** In existing valley physics, opposite Berry curvatures are located at different valleys, resulting in appealing phenomena, such as valley-polarized transport. Whereas, in our elastic valley metamaterials, the observed anisotropic band structure may result in the exceptional topology of valley, which has never been explored. We study the topology of valley by calculating the Berry curvature $\Omega(\boldsymbol{k}) = i\nabla_{\boldsymbol{k}} \times \langle u(\boldsymbol{k})|\nabla_{\boldsymbol{k}}|u(\boldsymbol{k})\rangle$ and integrate it using the discrete integration method (see Supplementary Note 2)[29]. For an elastic system with time reversal symmetry, the integration of the Berry curvature of the whole Brillouin zone is expected to be zero. Nevertheless, the Berry curvature is highly localized at the valleys and the local integration of the Berry curvature converges quickly to a nonzero quantized value. The global integration is referred to as the Chern number *C*, and the local one is called valley Chern number $C_v$. The valley Chern number is defined as $C_v = \frac{1}{2\pi} \int \Omega(\boldsymbol{k}) d^2\boldsymbol{k}$, where the integral bounds extend to a local area around the valley. In existing valley physics, the valley Chern numbers were calculated to be ±1/2 in electronic, photonic and phononic systems. Usually, the extrema of the Berry curvature are located at the corners of Brillouin zone. However, the maximum and the minimum values of the Berry curvature in our system are not technically at the corners of Brillouin zone. This discrepancy may be caused by mismatch between the asymmetrical spiral and the triangular lattice. Fig. 2a shows the regular Brillouin zone (Black solid lines) and the area formed by connecting the extrema of the Berry curvature (Red dashed lines). The area is clearly distorted but still has the symmetry around an oblique line (Blue dotted lines). The detailed Berry curvatures around each K valley are presented in Fig. 2b. The Berry curvature of the second band and the third band are enclosed by the cyan and magenta dotted lines, respectively. However, numerical integration of the Berry curvature of our system shows that the anomalous integration numbers are about ±0.34 that are not limited to ±0.5. This may be caused by the strong spatial inversion symmetry breaking[9,30]. The strong spatial inversion symmetry breaking reflects the distribution of slowness curves and group velocity, which is related to the Berry curvature by $\Omega(q) = \pm \frac{mv_g^2}{2(|q|^2v_g^2+m^2)^{\frac{3}{2}}}$, where $q = k - k_K$ is the wave vector deviation from the corresponding K point[24]. We also characterize the anisotropy of metamaterials by calculating the slowness curves and group velocity seen in the Supplementary Note 3 and Supplementary Figure 2. The distorted area and three different distributions of the Berry curvature around $K_1$, $K_2$, and $K_3$ valleys clearly show the anisotropic characteristic in the elastic valleys. The observed deviated distribution of the Berry curvature in elastic system may inspire the similar anomalous physics in other systems, such as electronic and photonic systems.

**The tunability of Berry curvature.** The Berry curvature plays an important role in wave motion in metamaterials and other fields in modern condensed matter physics. Thus, it is desirable to have more choices of the Berry curvature. We investigate the tunability of the Berry curvature, as a function of the key parameters of the Archimedean spiral structure, such as the number of turns (*n*) and the thickness of the spiral (*d*) as shown in Fig. 1a. For simplicity, we mainly investigate the Berry curvature along $K_6K_1$ line. The situation of the second band is shown in Fig. 3a. At *n* = 1.5 and 2 considered in this study, with the increase of the thickness of the Archimedean spiral, the absolution value of the Berry curvature decreases. It is quite understandable that since the Archimedean spiral approaches a circle with the increase of the thickness, the Berry curvature tends to disappear due to the spatial inversion symmetry. As the thickness of the Archimedean spiral is fixed, the sign of the Berry curvature changes when the number of turns changes from *n* = 1.5 to *n* = 2.0, where the peak sign flips, and thus the topological transition happens. The same phenomenon can be observed from Fig. 3b which shows the distributions of the Berry curvature of the third band.

**Elastic topological valley Hall edge states.** We have demonstrated the valley of the anisotropic elastic metamaterials in band structures, the deviated distributions of the Berry curvature, and its tunability. Naturally, the next step is to investigate the elastic topological valley Hall edge states, which is regarded as one of the most significant manifestations of valley-polarized behaviors. According to the result in the previous section, we can tune the value of the Berry curvature around valleys by adjusting number of turns ($n$) and thickness ($d$). We use the symbol ($n_1$, $d_1$ | $n_2$, $d_2$) for convenience to discuss the combination of different configurations of elastic valley metamaterials. We choose the combination of (1.5, 3) and (2, 2) for the investigation of elastic topological valley Hall edge states due to their overlapped frequency range and opposite Berry curvatures. As shown in Fig. 4a, the projected band along the $k_x$ direction is calculated using the sandwich supercell (2, 2 |1.5, 3 |2, 2). It is evident that there are three bands independent of the bulk shown in gray. The red line represents the edge states located at the interface between (2, 2) and (1.5, 3), while the blue line represents the edge states located at the interface between (1.5, 3) and (2, 2).

The projected band structure along the $k_y$ direction is displayed in Fig. 4b. Red and blue lines have the same meanings as those in Fig. 4a. However, when checking the eigenmodes of edge bands along the $k_x$ and $k_y$ directions, we find that as $k$ approaches the origin point from the intersection point, the edge states fade gradually in the first band along the $k_x$ direction and in the both bands along the $k_y$ direction. Although the displacement field of the first band in Fig. 4a mainly concentrates at the interface, its intensity concentrates on the spiral which looks more like bulk states of spirals, which is nearly non-dispersive. The detailed eigenmodes are shown in supplementary figure 3. This phenomenon of the branch fading is distinct from the common sense of valley topological bands where the robustness of the edge modes remains basically the same as $k$ varies. This can result in several intriguing phenomena of valley edge transport of electrons, photons and phonons. Therefore, the unique transport of elastic waves would be expected in our system when the direction of transport changes from $x$ direction to $y$ direction.

Here, we demonstrate a frequency selector using this novel property in our system. Fig. 4c and 4d show the schematics and the simulation models. When the vibration source with 157.6 Hz marked in cyan line in Fig. 4a and 4b, is set at the beginning of the interface between (2, 2) and

(1.5, 3), the elastic waves will transport along the interface. However, when the elastic waves arrive at the intersection, they will not go forward but go downward into another interface. The reason is evident: edge modes in different interfaces belong to different valleys and they cannot couple with each other, so the excited edge mode cannot propagate forward. But the downward edge mode and the excited mode are projected by the same valley. Therefore, the elastic waves go downward when they arrive at the intersection[17].

However, when the excitation frequency is low, the transport of the elastic waves is distinct from the existing cases. Here, we choose 156.7 Hz marked in magenta in Fig. 4a and 4b as the excitation frequency. When the elastic waves arrive at the intersection, they will neither go forward nor bisecting into another interface. From the projected band structure along the $k_x$ and $k_y$ directions, we know that the edge modes shown in red line in Fig. 4a can be excited. However, when the direction is changed to the $k_y$ direction, the edge mode cannot support the energy along the vertical interface (see the eigen displacement field in supplementary figure 3). Therefore, the elastic waves will not travel along the bent interface. The topological state attenuation with wave vector displays the unique wave dynamics, which has never been realized in other systems yet. It may have potential applications in signal processing and frequency selector.

**Discussion**

As a conclusion, the valley anisotropy is introduced by asymmetrical elastic metamaterials made of bio-inspired structure and soft material. The observed anomalous valley vortex states enrich the intrinsic physics of the valley states. The deviated Berry curvature in elastic system may inspire similar explorations in other systems, such as electronic and photonic systems. Note that the discussed elastic waves with valley anisotropic characteristics are transverse waves, which opens a new DOF to manipulate the transversal polarized waves. The hydrogel-based transverse wave manipulator can be glued to a variety of media including hard and soft materials due to its strong chemical anchorage effect[31]. Given its biocompatibility, it would be beneficial to use hydrogel-based devices in contacting with the skin. The novel manipulation of transverse wave has potential applications in biomedical field for elastography and in nondestructive flaw detection such as detecting the defect in metals and composites.

**Methods**

**Simulations.** Numerical simulations are performed by using COMSOL Multiphysics, finite-element analysis and solver software. The simulations are implemented in the 2D solid mechanics module. The system consists of the rigid spiral resonator made of polylactic acid (PLA) and soft hydrogels. The mechanical properties used for spiral PLA are: mass density 1250 kg·m$^{-3}$, Young's modulus 2.1 GPa and Poisson's ratio 0.36. The geometric parameters of the spiral are defined in Fig. 1a, in which initial spiral radius $a_i$ = 1.5 mm, gap distance $g$ = 2.25 mm, thickness of the spiral $d$ = 2 mm and the number of turns $n$ = 2. The mechanical properties of soft hydrogels are mass density 1000 kg·m$^{-3}$, Young's modulus 18 kPa and Poisson ratio 0.5. The lattice constant is $c$ = $14\sqrt{3}$ mm. For calculating the Berry curvature, the first-principle discrete method is used (see Supplementary information). For the calculation of wave propagation in a finite sample, a vibration source is set at the beginning of the interface and the absorption boundary conditions are set around the sample.

## Data availability

The data sets generated and analyzed during the current study are available from the corresponding authors on reasonable request.

## Acknowledgements

We gratefully thank Prof. Meng Xiao from Wuhan University, Prof. Feng Li from South China University of Technology and Dr. Rajesh Chaunsali from Laboratoire d'Acoustique de I'Universite du Maine for fruitful discussions. S.L. and J.Y. are grateful for the support from NSF (CAREER1553202 and EFRI-1741685). J.Z. are grateful for the support from the National Key Research and Development Program of China (2018YFB1105100) and the National Natural Science Foundation of China (51572096 and 51820105008). I. K. and S. I. are grateful for the support of MEXT KAKENHI Grant Number JP17J09077, JP15H05700, JP17H06138, JP15H05868.

**Author contributions**

S.L. did the simulations and wrote the manuscript. I.K. assisted with the calculation of the Berry curvature. S.I., J.Z., and J.Y. supervised this project. All authors were involved in analysis and discussion of the results and the improvement of the manuscript.

**Additional information**

**Competing financial interests:** The authors declare no competing financial interests.

**Corresponding author:** correspondence to Jianfeng Zang and Jinkyu Yang

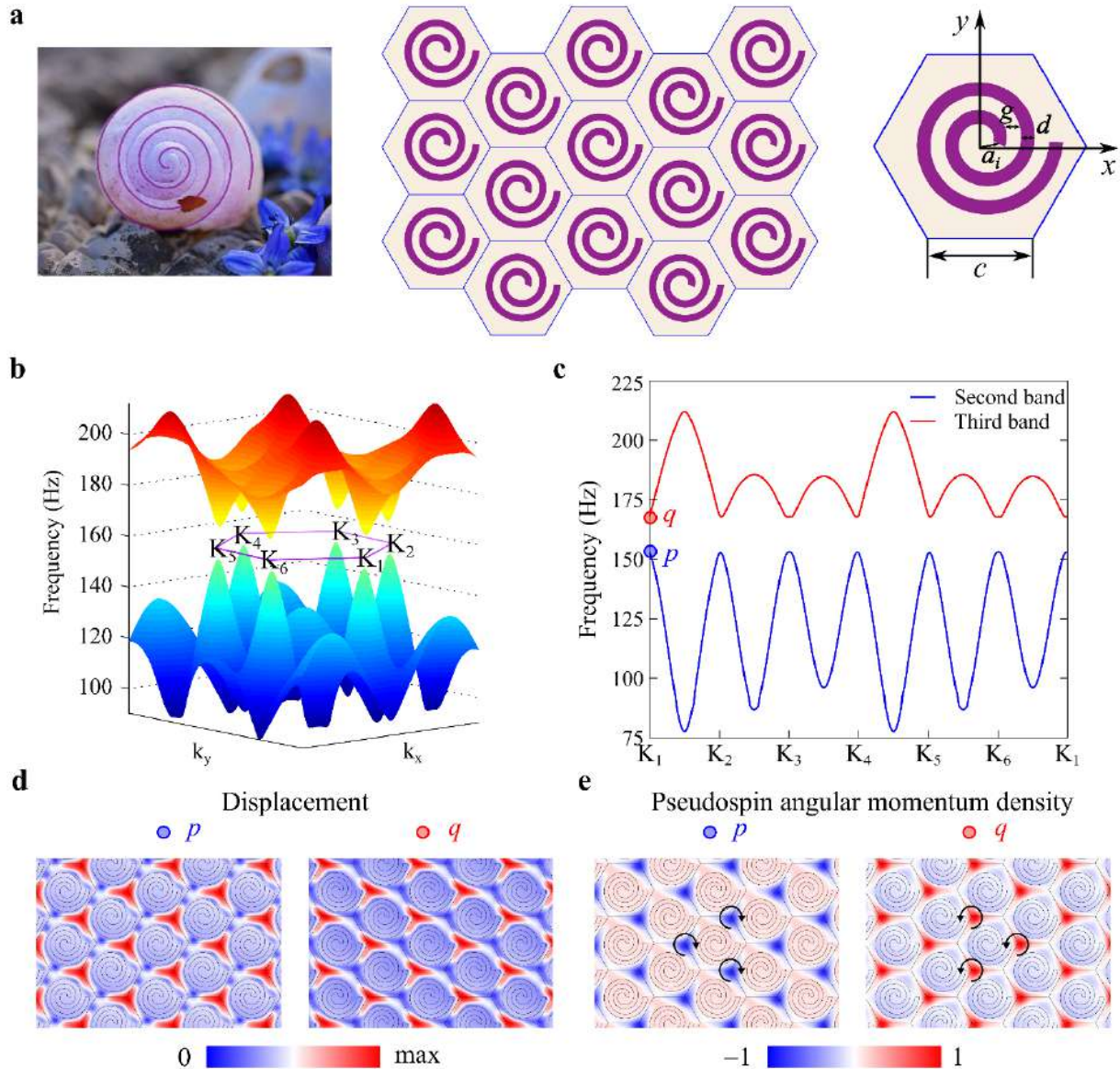

**Fig. 1 | The design of elastic valley metamaterials. a.** Schematics of the elastic valley metamaterial in triangle lattice. The Archimedean spiral-like structure inspired by seashells is employed as the rigid scatterer shown in purple. The soft hydrogel matrix is shown in beige. **b.** The band structure of the first Brillouin zone with the second band and the third band showing the valley characteristic. The purple lines indicate the first Brillouin zone and the six corners are marked by $K_1$, $K_2$, $K_3$, $K_4$, $K_5$ and $K_6$. **c.** The band structure along the edge of the first Brillouin zone. The blue (red) curve represents the second (third) band. The two states in the $K_1$ valley denoted as $p$ and $q$ states. **d.** The displacement field distributions of $p$ and $q$ valley states. **e.** The normalized pseudospin angular momentum density distributions of $p$ and $q$ valley states.

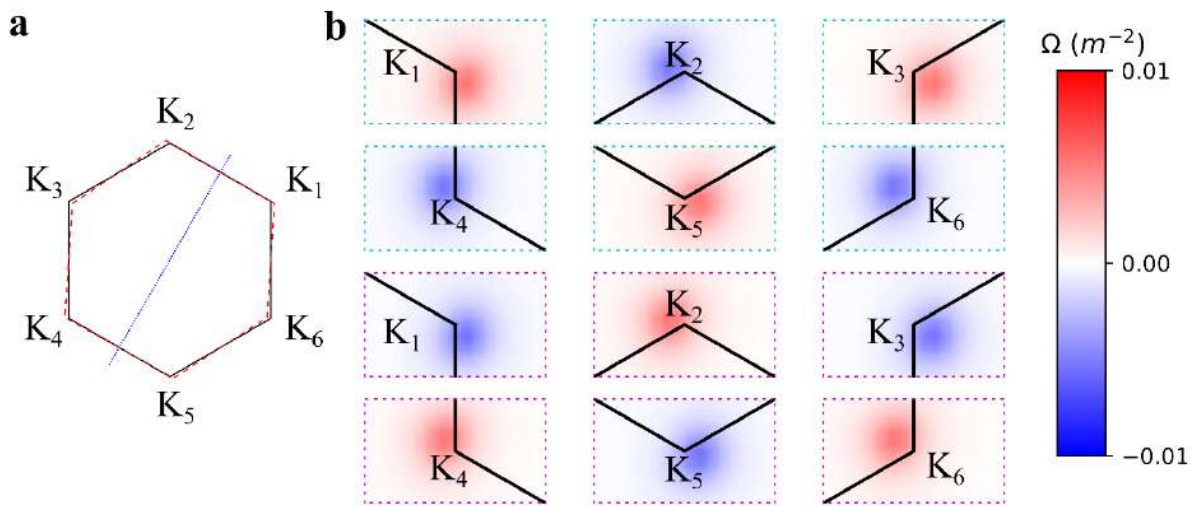

**Fig. 2 | The Berry curvature near K points of the second band and the third band. a.** The black solid lines represent the Brillouin zone and the red dashed lines connect the extrema of the Berry curvature. The blue dotted line represents the symmetrical plane. **b.** The detailed Berry curvature distribution near the six valleys enclosed by the cyan dotted line (The second band) and magenta dotted line (The third band). The black lines represent the boundaries of Brillouin zone.

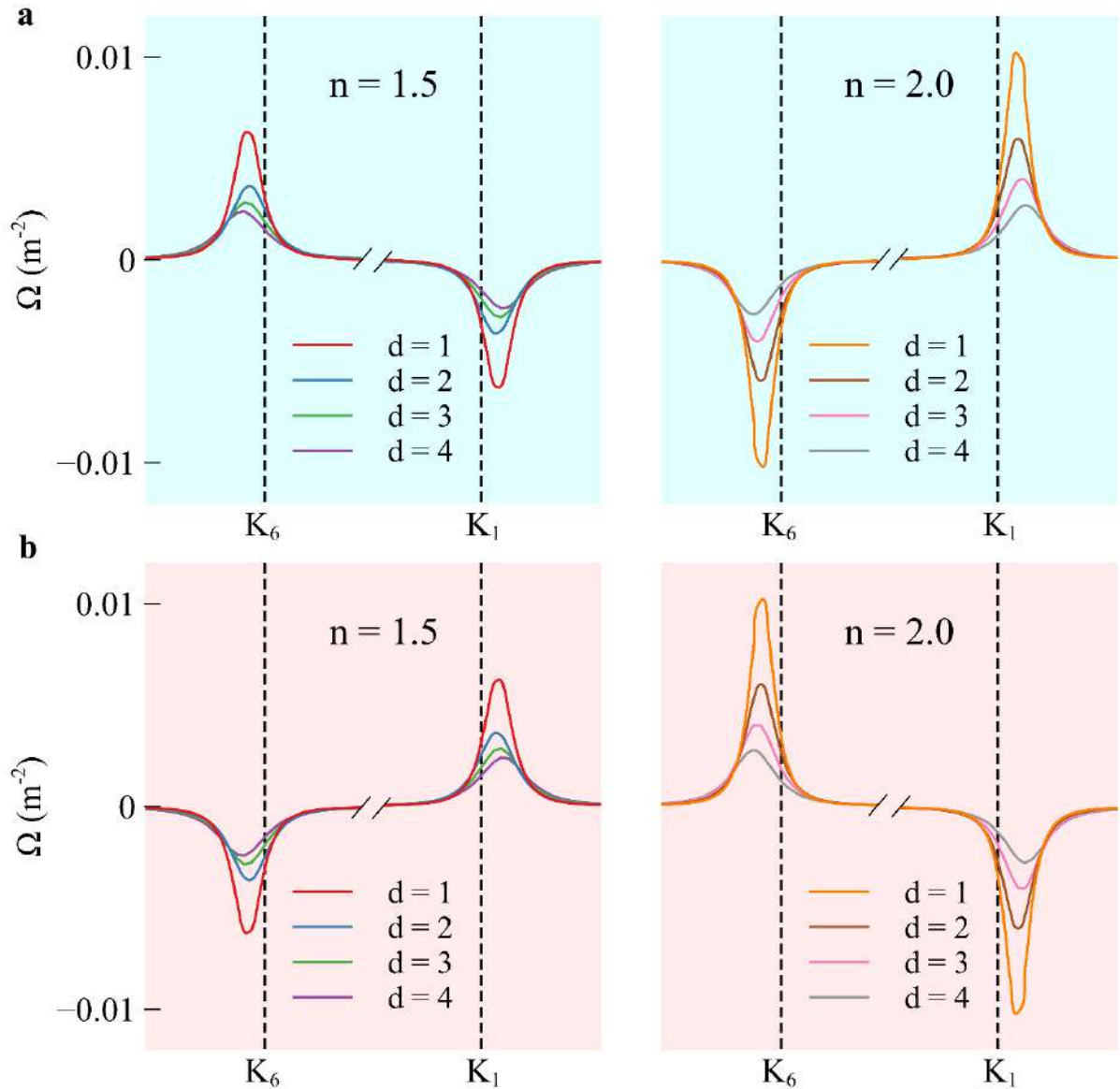

**Fig. 3 | The tunability of the Berry curvature. a.** The Berry curvature of the second band along $K_6 K_1$ line as a function of the thickness of the Archimedean spiral, $d$ = 1, 2, 3 and 4, when the number of turns of the spiral $n$ = 1.5, 2. **b.** The Berry curvature of the third band along $K_6 K_1$ line as a function of the thickness of the Archimedean spiral, $d$ = 1, 2, 3 and 4, when the number of turns of the spiral $n$ = 1.5, 2. Note that there is a broken *x* axis in each figure to emphasize the change of the Berry curvature in the K points.

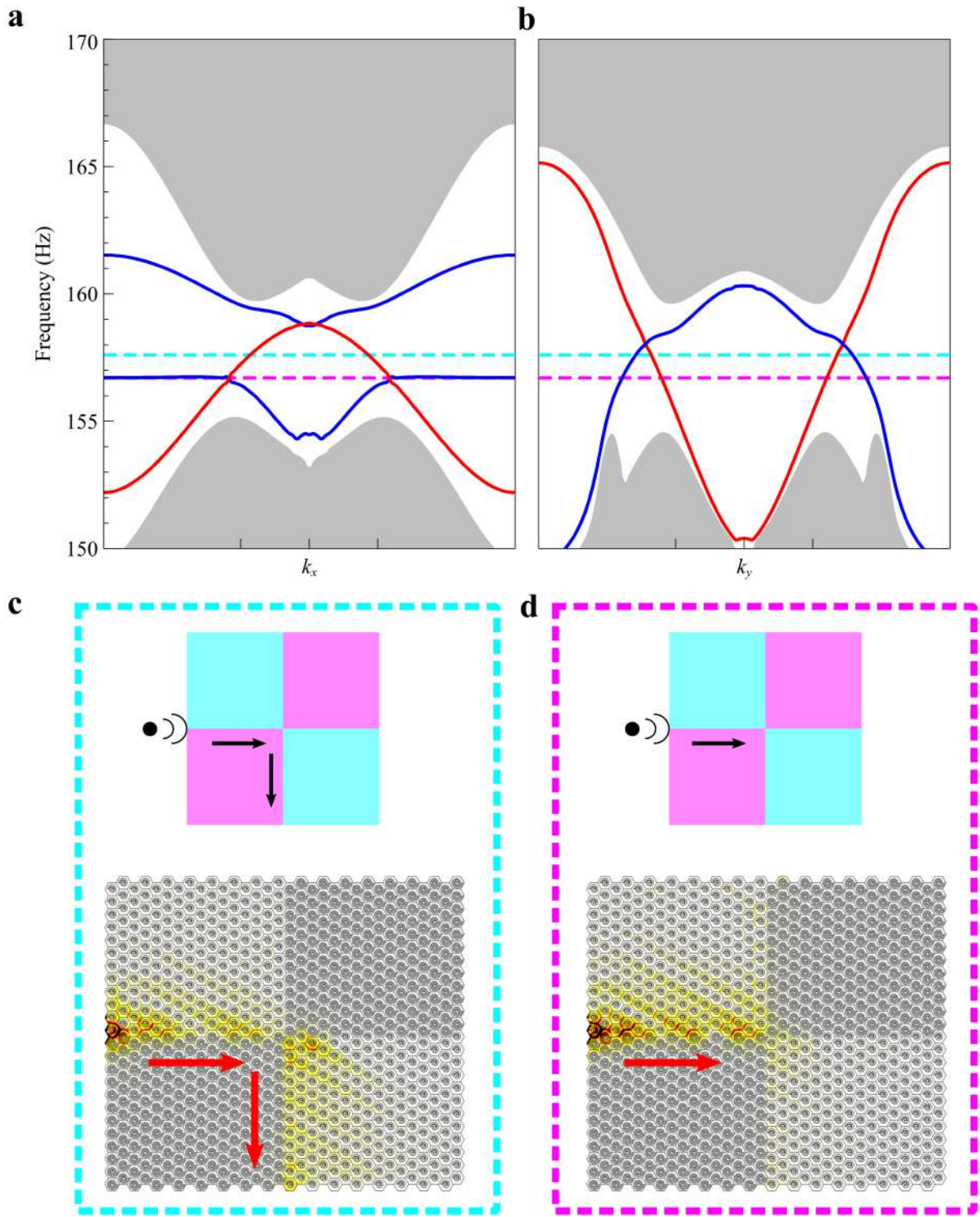

**Fig. 4 | The elastic topological valley Hall edge states a.** The projected band structure along $k_x$ direction with two distinct modes shown in red and blue solid line. The bulk modes are shaded in gray. **b.** The projected band structure along $k_y$ direction with two distinct modes shown in red and

blue solid line. The bulk modes are shaded in gray. **c.** The transport of elastic wave along the interface. At the frequency of 157.6 Hz, which is shown in cyan line in Fig. **a** and **b**, the elastic wave can travel along the path and through the bend. **d.** The transport of elastic wave along the interface. At the frequency of 156.7 Hz, which is shown in magenta line in Fig. **a** and **b**, the elastic wave can travel along the $x$ direction but cannot travel through the bend. In schematics of both **c** and **d**, the cyan blocks represent (1.5, 3) metamaterials and the magenta blocks represent (2, 2) metamaterials, which indicate the simulation setup below.

# Supplementary Information

# Valley Anisotropy in Elastic Metamaterials

**Shuaifeng Li[1,2], Ingi Kim[3], Satoshi Iwamoto[3], Jianfeng Zang[2*], Jinkyu Yang[1*]**

[1]Aeronautics and Astronautics, University of Washington, Seattle, Washington, 98195-2400, USA

[2]School of Optical and Electronic Information and Wuhan National Laboratory for Optoelectronics, Huazhong University of Science and Technology, Wuhan, Hubei, 430074, China

[3]Institute of Industrial Science, University of Tokyo, Meguro, Tokyo, 153-8505, Japan

## Supplementary note 1

**Comparison between traditional acoustic valley metamaterials and our metamaterials**

Here, we use the one of the most widely studied structures that achieve acoustic valley vortex states to make a comparison with our system[1–3]. We construct the acoustic metamaterial consisting of an equilateral triangular rod in the water with triangular lattice. The equilateral triangular rod is made of steel with $\rho = 7670\ kg/m^3$, $c_p = 6010\ m/s$, $c_s = 3230\ m/s$ and the acoustic parameters for water are $\rho = 1000\ kg/m^3$, $c_p = 1490\ m/s$. The lattice constant is $14\sqrt{3}\ mm$ and the side of equilateral triangular rod is $20\ mm$. When the symmetries between the lattice and scatterer match, the Dirac cone is expected to appear at K point. In order to open the Dirac cone, we need to break the spatial inversion symmetry by rotating the steel rod. Here the rotation angle is 10° as shown in the inset of supplementary figure 1.

Supplementary figure 1a shows the band structure of this acoustic metamaterial. The Brillouin zone and the high symmetrical points are shown in the inset. Clearly, there is a band gap around 30 kHz, and the band structure shows two extrema at K point, which are denoted as *p* and *q* states respectively. To show the valley features, we investigate the eigen pressure field and pseudospin angular momentum density distribution similar to what we have done in our main text. As displayed in supplementary figure 1b, when the eigen pressure fields reach maximum in the corners of hexagon in both *p* state and *q* state, the pseudospin angular momentum densities tend to be zero. We also notice that the *p* state is in the pseudospin up state and the local particle velocity rotates counterclockwise while *q* state is the opposite. In sharp contrast, in our system, when the eigen fields reach maximum in the corners of the hexagon in both *p* and *q* states, the amplitudes of pseudospin angular momentum densities also tend to be maximum.

## Supplementary note 2

**Calculation of Berry curvature**

After obtaining the dispersion relation ω = ω(**k**) and displacement vector field **U**(**k**) through finite element method (FEM), we calculate the Berry curvature by numerical method[4]. For our two-dimensional system, we consider a clockwise path around a certain point A ($k_x$, $k_y$) consisting of $A_1$ ($k_x$-$\delta k_x$/2, $k_y$-$\delta k_y$/2), $A_2$ ($k_x$-$\delta k_x$/2, $k_y$+$\delta k_y$/2), $A_3$ ($k_x$+$\delta k_x$/2, $k_y$+$\delta k_y$/2) and $A_4$ ($k_x$+$\delta k_x$/2, $k_y$-$\delta k_y$/2). According to the definition and Stokes' theorem, we obtain $\int \Omega \, d^2 k = -\int \boldsymbol{B} \cdot d\boldsymbol{k}$, where **B** is the Berry potential of a state defined by $i\langle U_k | \nabla_k U_k \rangle$.

Since we consider the continuous Brillouin zone as numerous small patches, for each patch $\delta k_x \times \delta k_y$, we estimate the Berry curvature as below:

$$\Omega(A) = \frac{\text{Im}[\langle U(A_1)|U(A_2)\rangle + \langle U(A_2)|U(A_3)\rangle + \langle U(A_3)|U(A_4)\rangle + \langle U(A_4)|U(A_1)\rangle]}{\delta k_x \times \delta k_y}$$

where the inner product can be calculated in Comsol. Then, we can map the Berry curvature of the Brillion zone shown in Fig. 2. Further, the local integral is often referred to as the valley Chern number $C_v$ of the *n*th band and it is defined as $C_v = \frac{1}{2\pi} \int \Omega(\boldsymbol{k}) d^2 \boldsymbol{k}$.

## Supplementary note 3

**The anisotropy of elastic valley metamaterials**

To better understand the anisotropic nature of the elastic valley metamaterials, we further calculate the slowness curves and group velocity distribution. The slowness curves and group velocities are evaluated as a function of the wave vectors. Supplementary Figure. 2a and 2b present the polar plots of the slowness curves of the second band and the third band, showing the distribution of slowness magnitude for different directions in the first Brillouin zone. The slowness curves corresponding to the second and third band are approximately circular at small wave vectors. Interestingly, the slowness curves exhibit the evident anisotropy as wave vectors increase. As wave vectors increase, the values of the slowness of the second band and the third band rise, which is consistent with the corresponding band structures in Fig. 1b. As the wave vector approaches the edges of the first Brillouin zone, the slopes of the corresponding bands decrease. It is noted that the slowness curves are symmetrical around the central point, which agrees with the band structures in Fig. 1c.

The calculated group velocity distribution of the second and the third band are presented in Supplementary Figure. 2c and 2d, respectively. It can be found that the group velocity patterns of both bands are quite complex, especially at the corners of the Brillouin zone. There is a single point with nearly zero group velocity around each K point. The group velocity distribution of both bands as a function of wave vectors is symmetrical around the center of the Brillouin zone, suggesting the chiral anisotropic structure of the elastic valley metamaterials, which is correspondence of the slowness curves

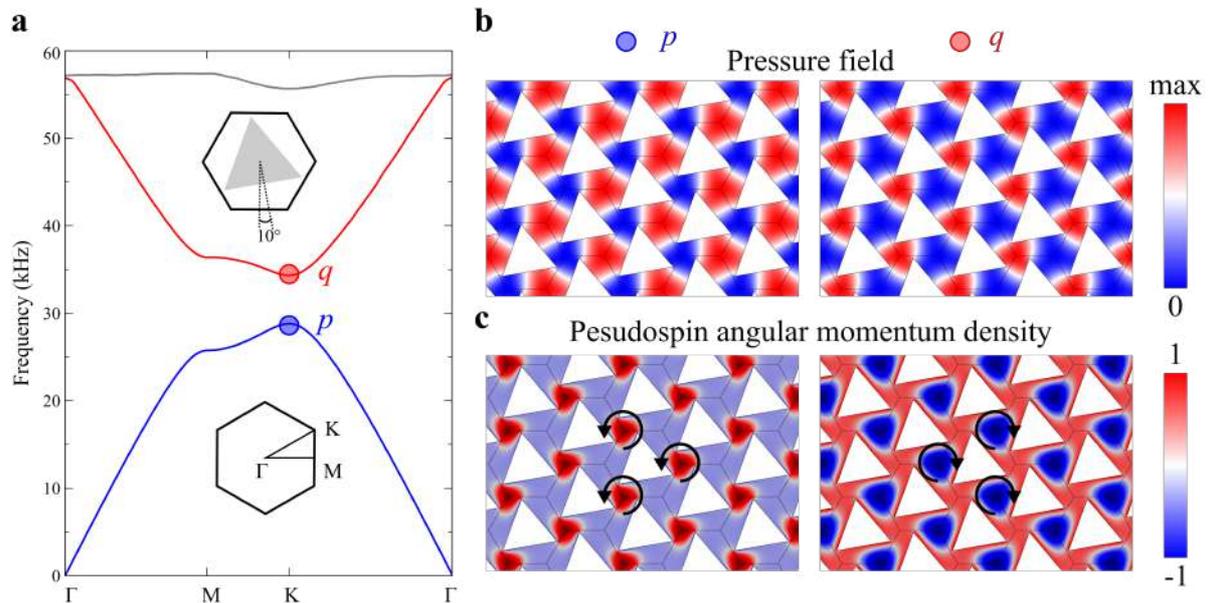

**Supplementary Figure 1 | The comparison of traditional acoustic valley metamaterials. a.** The band structure along Γ-M-K-Γ of the acoustic valley metamaterial for comparison. The blue, red and gray curves represent the first, second and third band, respectively. **b.** The amplitude distributions of pressure fields of *p* and *q* states. **c.** The normalized pseudospin angular momentum density of *p* and *q* states.

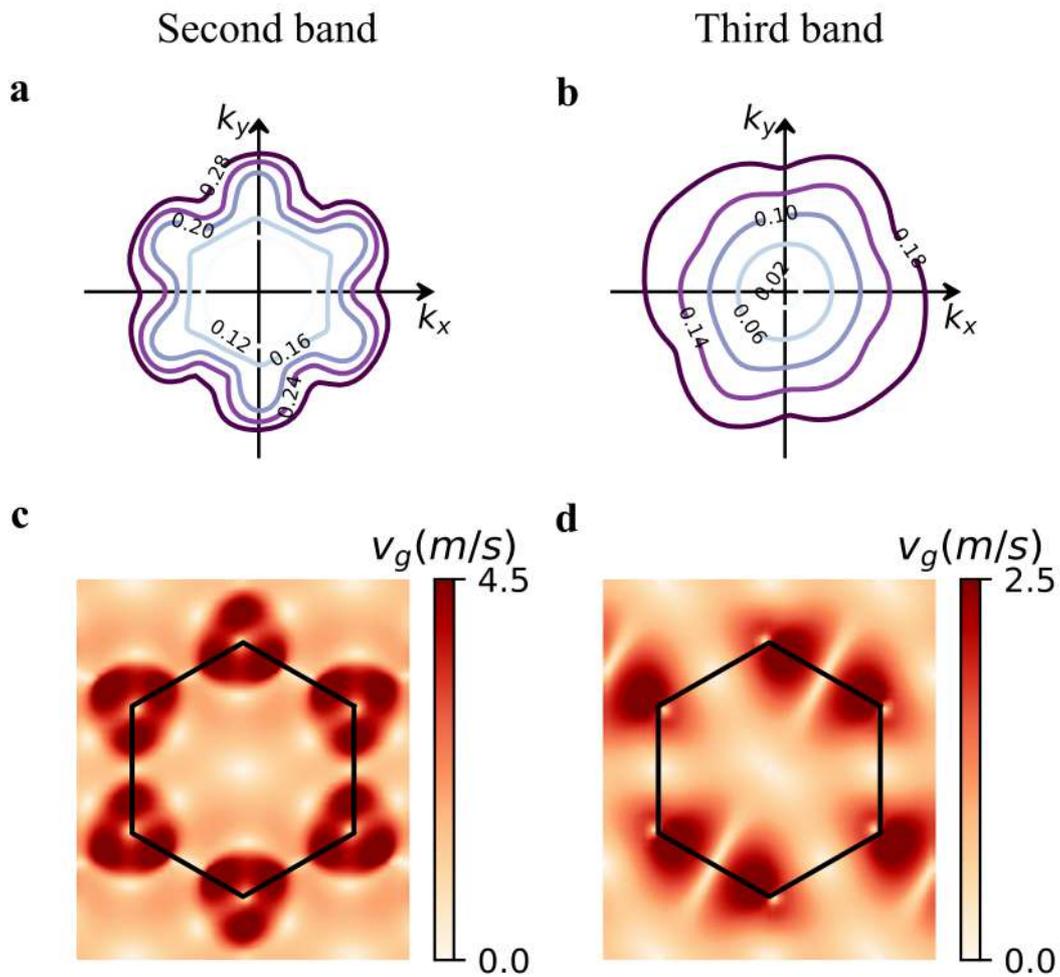

**Supplementary Figure 2 | The chiral anisotropy characteristics of the elastic valley metamaterials. a.** The slowness curves of the second band. From the small wave vectors to large wave vectors, the slowness rises from 0.12 to 0.28. **b.** The slowness curves of the third band. From the small wave vectors to large wave vectors, the slowness rises from 0.02 to 0.18. **c.** The group velocity distribution of the second band. The black solid line indicates the Brillouin zone. **d.** The group velocity distribution of the third band. The black solid line indicates the Brillouin zone.

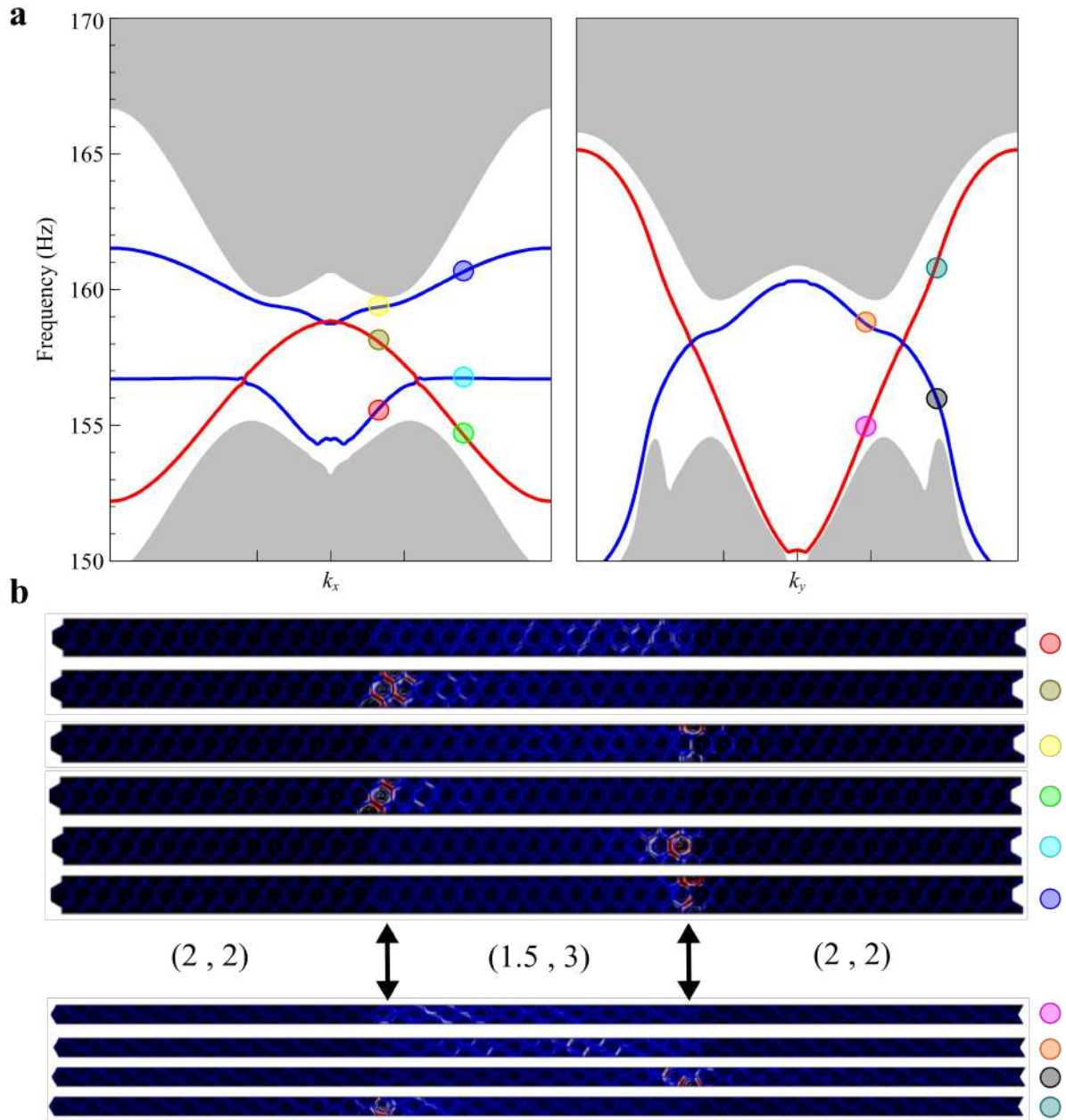

**Supplementary Figure 3 | The projected band structures and eigenmodes. a.** The projected band structures along the $k_x$ (left panel) and $k_y$ (right panel) directions. Several markers with different colors are shown in the topological band. For the projected band structure along the $k_x$ direction, markers are placed on each topological band at $k_x = 0.26$ and $k_x = 0.6$. For the projected band structure along the $k_y$ direction, markers are placed on each topological band at $k_y = 0.3$ and $k_y = 0.6$. **b.** The eigen displacement fields corresponding to the markers indicated. The setup of sandwich supercell is shown by the notation we mention in the main text and the arrows indicate the positions of interfaces.